# A Cloud-based Service for Real-Time Performance Evaluation of NoSQL Databases

Omar Almootassem, Syed Hamza Husain, Denesh Parthipan, Qusay H. Mahmoud
Department of Electrical, Computer, and Software Engineering
University of Ontario Institute of Technology
Oshawa, ON, L1H 7K4 Canada

*Abstract*—We have created a cloud-based service that allows the end users to run tests on multiple different databases to find which databases are most suitable for their project. From our research, we could not find another application that enables the user to test several databases to gauge the difference between them. This application allows the user to choose which type of test to perform and which databases to target. The application also displays the results of different tests that were run by other users previously. There is also a map to show the location where all the tests are run to give the user an estimate of the location. Unlike the orthodox static tests and reports conducted to evaluate NoSQL databases, we have created a web application to run and analyze these tests in real time. This web application evaluates the performance of several NoSQL databases. The databases covered are MongoDB, DynamoDB, CouchDB, and Firebase. The web service is accessible from: *nosqldb.nextproject.ca*.

## I. INTRODUCTION

The collection, storage and retrieval of data is the fundamental aspect of the digital world. As time has progressed, the ability for computers to process data has increased exponentially. "Experts say the world's data is doubling every two years" [5]. The relational database model was and is still adequate to handle most kinds of data sets however; the surge of big data, the need for high availability and the current application development methods led to the development of NoSQL databases.

NoSQL databases provide various features that allow it to handle large sets of complex, random and unstructured data. Some of the characteristics of NoSQL databases are: schema less design, non-relational structure, high scalability and it is highly distributable [5]. NoSQL databases consist of four types: key-value, document-based, column-based and graph-based. Each database type performs variably against common attributes such as performance, scalability, complexity and functionality that make it uniquely advantageous. Among the four database types, numerous databases can be selected based on the priority of the desired attributes.

The purpose of this paper is to present the design and implementation of a cloud-based service for real-time performance evaluation of NoSQL databases. Database performance evaluation allows the user to make an informed decision on which database to use for their application. "The choice of a particular NoSQL database imposes a specific distributed software architecture and data model, and is a major determinant of the overall system throughput" [10]. Although there are several NoSQL databases, the four most commonly used ones CouchDB, DynamoDB, MongoDB and Firebase were chosen for evaluation.

In this paper, we provide a background to the related work of the project. This is done to offer insight to previous performance analysis methods and to help guide the development of our own test cases. We then propose a solution by developing a web application to allow users to evaluate any of the four databases against our test cases. We discuss the results of our analysis and evaluate the strengths and weaknesses of the databases. We conclude by, describing what our application accomplishes and suggest any future means to enhance it. To this end the contribution of this paper is a cloud-based service for real-time performance evaluation of NoSQL databases, with evaluation results from specific configurations as discussed.

The rest of the paper is organized as follows. Section II discusses the background concepts of NoSQl databases and the related work. Section III presents the design and implementation of the cloud-based service. Evaluation results are presented in Section IV, and challenges and solutions are discussed in Section V. Section VI concludes the paper and provides ideas for future work.

## II. BACKGROUND AND RELATED WORK

There are several reports written based on evaluation of NoSQL databases. The evaluation is based on the particular choices of test cases chosen and the databases they are performed on. Among the numerous evaluations done on the databases we've covered, there are a common group of test cases used:
- Read and Update Small Data operations
- Retrieve Large Data operations
- Read and Update Large Data operations
- Read and Insert Data operations
- Consists of randomly distributed read and complex modify/write operations
- Insert Large Data operations

Unlike the orthodox static tests and reports conducted to evaluate NoSQL databases, we have created a web application to run and analyze these tests in real time. This web application evaluates the performance of several NoSQL databases. The databases covered include but are not limited to MongoDB, DynamoDB, CouchDB, and Firebase. The testing process consists of identification and creation of several test cases that are performed on the databases. The evaluation consists of determining which database optimizes the specified test case. An analysis of the results is used to determine the applicability of the database depending on the situation. Our test cases primarily cover:

- Uploading Small Data
- Uploading Large Data
- Retrieving Small Data
- Retrieving Large Data
- Updating Small Data
- Updating Large Data

### III. CLOUD-BASED SERVICE

As the topic involved testing numerous NoSQL databases, we decided to build a cloud-based service using NodeJS to provide a testing platform for the end user. The user can also examine the test results from other users.

*A. Assumptions*

In the infant stages of development, we needed to derive some criteria for evaluating the performance of the NoSQL database. From our research of common NoSQL functionalities, we have identified that uploading, retrieving, and updating are the most commonly used functions. We assume that for our current evaluation, those will be the 3 that will be tested on each database. Moreover, we want to test two different types of data sets to see how well each database performs for each function. We assume that 200kb is sufficient to reflect a large data file, and 5kb to reflect a small data file.

*B. Architecture*

What we have built is a full stack application with a front end built using single page architecture. Both the frontend and the backend are built using NodeJS which uses full JavaScript architecture. There are 3 major components to this application; the client, the application server, and the database servers.

The client is built using AngularJS framework and Angular Material UI framework. We chose to use a UI framework to keep our focus on implementing the functionality instead of spending a lot of time to make the application look attractive. We used some modules as well to help with creating the application. These module include md-steppers for the test tab, angular-timer to keep track of how long each test takes to complete and angular-chart to display the data in a visually appealing way. Express was also installed to deploy the application to EC2 on Amazon Web Services. The client connects directly to Firebase [2] instead of going on the server to execute the calls.

The application server is a NodeJS server built using jade templating engine. ExpressJS is used to create routes to communicate with the databases. Mongoose [8] is used to connect to MongoDB, Cradle [4] connects to CouchDB and aws-sdk is needed to connect to DynamoDB [1]. The application server is hosted on the same instance as the client, but is uses another port in the virtual machine. We also used cors module to allow cross origin resource sharing between the client and the server. The default setup of the server allowed uploading data less than 5MB but some of our datasets are larger than that so we increased the limit to 50MB.

The database servers are hosted on Compute Engine on Google Cloud Platform. The database servers are used for MongoDB and CouchDB since they are designed to be installed on a machine and not just run from a cloud service. The images we used are pre-configured virtual machines provided by Bitnami. Firebase and DynamoDB do not need servers to run as they just run from their own respective cloud service.

*C. Implementation*

Despite having 3 architectural components, the database servers were already set up and the correct ports were open so we didn't need to do any setting up on those servers. Most of the work on the application was done on the client. The client is what the user sees, so we focused on ensuring everything is functional and is very easy to use and understand. There are 2 major components for the client. The testing component and the charts and visualization component.

The testing components runs the test on the databases and saves the results in Firebase. We have 6 tests which are: uploading small and large amounts of data, retrieving small and large amounts of data, and updating small and large amounts of data. There are four databases that can be used to test which are MongoDB, Firebase, DynamoDB, and CouchDB. The system is designed to allow the user to run one test at a time on as many databases as the user wants. For example, the user can retrieve small amounts of data on Firebase and DynamoDB at the same time but cannot upload small amounts of data and update large amounts of data on Firebase at the same time. After performing the tests, the time taken is saved in the Firebase database as well as the user's location which comes from freegeoip.net.

The chart and visualization component is what the user is greeted with when they navigate to the website. The main chart on that page displays the average time taken to perform each test on each database regardless of location. Below that chart, there are 4 smaller charts display the best and worst scenarios for each tables performing all the tests. These charts display real time data and are automatically updated after any test is performed, so the user will always have the most recent data available. The last thing on the page is Google Map displaying a heat map of where the tests were taken. This is added to give the user an idea of where the previous tests were performed from.

The server uses ExpressJS to create routes to perform REST methods such as PUSH, PATCH, PUT and GET. Each route refers to a different command and a different database. For example calling /post/mongo_data will post data to the MongoDB on Google Cloud Services. As stated previously in the architecture, we used mongoose to connect to the MongoDB instance on Google Cloud and Cradle to connect to the CouchDB [9] instance on Google Cloud. We used aws-sdk to connect to DynamoDB.

*D. Scalability*

The web service we have designed does not expect to have millions of users active at the same time, so scalability wasn't a big focus while developing the application. However, future proofing is always important in any project. Scaling means running multiple instances of the same image, so we do not want to have any information saved locally on the virtual machine. Our databases are hosted on a different server than our application. MongoDB and CouchDB are hosted on Compute Engine on Google Cloud Platform. DynamoDB [1] is on AWS and Firebase [2] has it's own container on Google Cloud Platform. With this setup, if the application receives an unexpected influx of users and is forced to scale up, the databases will not be affected.

The test results are stored in Firebase. Firebase has its own autoscaling setup to ensure that it does not crash or get overloaded. Firebase has a NoSQL database which is scalable. We also set up authentication for Firebase which happens behind the scenes when the application is loaded to protect the user's data.

## IV. EVALUATION RESULTS

After running the tests over 30 times from different locations, we came up with a few conclusions. Table I shows the average times (in millisecond or ms) taken to perform each test. Overall CouchDB takes the longest time to complete all of our tests. Uploading and updating large amounts of data takes over twice as much time to complete as the next slowest database, MongoDB.

Table I. AVERAGE TIME IN MS TAKEN TO PERFORM EACH TEST

|  | MongoDB | DynamoDB | Firebase | CouchDB |
| --- | --- | --- | --- | --- |
| Upload Small Data | 250 | 210 | 70 | 470 |
| Upload Large Data | 1200 | 680 | 500 | 2800 |
| Retrieve Small Data | 160 | 150 | 55 | 366 |
| Retrieve Large Data | 740 | 300 | 540 | 700 |
| Update Small Data | 250 | 210 | 40 | 520 |
| Update Large Data | 1280 | 680 | 380 | 2800 |

In terms of testing the best and worst case scenarios for each database performing each test, Firebase had the most stable performance out of all databases. The worst case test for firebase is uploading large data which took just over 1000 milliseconds.

CouchDB has good caching capabilities. Running a test multiple times in a row results in completing the test in zero milliseconds.

Table II. MAXIMUM TIME IN MS TO PERFORM EACH TEST

|  | MongoDB | DynamoDB | Firebase | CouchDB |
| --- | --- | --- | --- | --- |
| Upload Small Data | 290 | 270 | 179 | 666 |
| Upload Large Data | 2300 | 2150 | 1050 | 4800 |
| Retrieve Small Data | 170 | 230 | 110 | 400 |
| Retrieve Large Data | 1400 | 500 | 600 | 900 |
| Update Small Data | 250 | 220 | 70 | 700 |
| Update Large Data | 1800 | 1200 | 800 | 3300 |

From the data presented in Table II, we can conclude that CouchDB had the worst worst-case performance of any database we tested and Firebase has the best worst-case performance of the databases we tested.

DynamoDB and MongoDB had a similar performance as each other but DynamoDB has the edge when working with large data.

From Table III we see a few unexpected values. For example the best case scenario for CouchDB when uploading data is 0 milliseconds. This value only occurs when we upload the same data more than once without refreshing the test page, this is due the caching mechanism provided by CouchDB and Cradle. They check that the data being pushed is the same as the previous data and creates another reference for the previous data instead of re-uploading the same data over and over.

Ignoring the 0 millisecond values, Firebase has the best best-case performance out of all the databases we tested and CouchDB has the worst best-case performance. The main reason why Firebase performed greatly in our tests is because it is connected directly to the client, instead of being connected to the server. This means there is less middleware to go through to access the database. An active connection to firebase remains open the whole time while the application is running. This allows Firebase to bypass the connection phase when running tests.

Table III. MINIMUM TIME IN MS TO PERFORM EACH TEST

|  | MongoDB | DynamoDB | Firebase | CouchDB |
|---|---|---|---|---|
| Upload Small Data | 163 | 100 | 35 | 0 |
| Upload Large Data | 400 | 0 | 150 | 0 |
| Retrieve Small Data | 150 | 120 | 30 | 300 |
| Retrieve Large Data | 400 | 450 | 400 | 500 |
| Update Small Data | 220 | 200 | 30 | 400 |
| Update Large Data | 850 | 500 | 130 | 2600 |

## V. CHALLENGES AND SOLUTIONS

The biggest challenge building this application was researching which databases to use and finding reliable cloud hosting for the all the databases. Connecting the databases to the application also proved to be a challenge since every database has its own unique method to connect to it.

The simplest database to connect to was Firebase. Since Firebase [2] is a backend as a service, it did not require creating a server to connect to it. We could easily connect to it from the client by pasting a snippet of code provided in the Firebase console. Since our application is built using AngularJS framework, we had the option of using AngularFire to simplify the process even more, but we opted against it. AngularFire [3] provides AngularJS bindings for Firebase and is officially supported by Google. All in all connecting to Firebase was a breeze.

After the Firebase connection was established, we started connecting MongoDB. Unlike Firebase, MongoDB [7] will not run directly on the client; it needs a server. We set up a NodeJS server and used a node module called Mongoose to simplify the connection process. Mongoose [8] is an elegant MongoDB object modeling module for NodeJS. Since we are deploying the system as an auto scaling system, we cannot have MongoDB setup on the localhost of the system since scaling will cause duplication and loss of data. We looked into a few hosted services including MongoDB Atlas [6] but in the end we went with a MongoDB instance created by Bitnami and hosted on Google Cloud.

CouchDB had a similar setup process to MongoDB. We used a node module called Cradle [4] to simplify the connection process. The CouchDB instance is also hosted on Google Cloud and created by Bitnami.

DynamoDB had a different setup process than the other databases. DynamoDB has three different methods to set up a connection. All three required using the aws-sdk node module. We chose to use the simplest method which is creating a /.aws folder in the root directory of the system which contains 2 files, config and credentials. The 2 files contains the configuration information and the credentials of the connected AWS account. When migrating the application to an EC2 instance on AWS, we needed to recreate these files on the virtual machine as well to authenticate the connection.

The second challenge we encountered was finding a unique and informative way to display the data that is collected after performing the tests. Visualization of this data is imperative. After examining various tools for HTML5 data visualization, there were two contenders: D3 and Chart.js with a comparison shown in Table IV.

Table IV. COMPARISON BETWEEN D3 and CHART.JS

| D3 | Chart.js |
|---|---|
| + Highly customizable<br>+ Verty versatile<br>+ Capable of visualizing a vast amount of data | + Simple to use<br>+ Integrates well with JavaScript |
| - Very complicated to use<br>- Many features are outside the scope of this application<br>- Requires advanced understanding of HTML5 canvas | - Can be too simple, design elements lack customizability<br>- Designed to work with embedded JavaScript, needs a wrapper to work with AngularJS |

Ultimately, simplicity was chosen over customizability, and Chart.js was selected. Since Chart.js is not designed to work with AngularJS, we needed to a node module called angular-chart which has Angular style calls for Chart.js.

We are also collecting location data from the users' IP addresses. We needed a way to display this data that wasn't in a chart. We chose to create a heat map that shows where the users are running the tests from and the colour shows the average speed in that location. We used Google Maps API V3 to create the heat map since we have a lot of experience working with Google Maps APIs on both mobile and web. The big challenge in web is that the API does not natively work with the AngularJS architecture so we needed to use a node module called AngularJS Google Maps which ported the API calls to Angular services and directives.

Another challenge we had was to choose a geolocation API. We initially settled on using ip-api.com; however, while testing the application on campus, we realized that that service is completely blocked on campus, so we switched to freegioip.net which provides an equally simple service.

## VI. Conclusion And Future Work

Evolving from static databases, the evolution of NoSQL has taken the market by storm and provided a suitable replacement for database use and many other services. However, projects are unique in the way they are built and run, therefore the requirements for each project will dictate which database is best to use. Some databases put emphasis on data retrieval time on the database and some databases put emphasis on update time. To gauge which one is best for your system, an evaluation process is ran on all databases where test cases are ran on each database.

In this paper, we have presented the design and implementation of a web service that evaluates the performance of a variety of NoSQL databases. Unique test cases were selected to test the performance of each database to get a primary idea of which performs the best.

The main criteria used to evaluate each test case is time. You can select a combination of MongoDB, DynamoDB, Firebase, and Couch DB and run six tests on each database. Once the tests are finished running, you can view the results in the "Data Visualization" tab. For each test, the results are illustrated as graphs and you can see which database is best fit for your own application.

Currently, the application only supports 4 databases. In future work, we could connect more databases to the application evaluate a larger group of databases. Moreover, to achieve a more precise evaluation on the overall performance of each application, we will add more test cases to run. The additive test cases will reflect some of the other functionalities that NoSQL databases can perform.